\documentclass[aps,prx,twocolumn,superscriptaddress,nofootinbib,notitlepage,longbibliography]{revtex4-1}
\usepackage{times}
\usepackage{graphicx}
\usepackage{amsmath}
\usepackage{amstext}
\usepackage{amssymb}
\usepackage{xfrac}
\usepackage[colorlinks,citecolor=blue]{hyperref}
\usepackage{graphicx}
\usepackage{amsmath}
\usepackage{amstext}
\usepackage{amssymb}
\usepackage{amsfonts}
\usepackage{longtable,booktabs}
\usepackage{hyperref}
\usepackage{url}
\usepackage{subfigure}
\usepackage{dsfont}
\usepackage{booktabs}
\usepackage{amsbsy}
\usepackage{dcolumn}
\usepackage{amsthm}
\usepackage{bm}
\usepackage{esint}
\usepackage{multirow}
\usepackage{hyperref}
\usepackage{cleveref}
\usepackage{mathrsfs}
\usepackage{amsfonts}
\usepackage{amsbsy}
\usepackage{dcolumn}
\usepackage{bm}
\usepackage{multirow}
\usepackage{color}
\usepackage{extarrows}
\usepackage{datetime}
\usepackage{comment}
\usepackage[super]{nth}

\hypersetup{
    colorlinks=true,
    linkcolor=blue,
    filecolor=magenta,
        urlcolor=blue,
}

\newcommand{\comments}[1]{}

\newtheorem{statement}{Statement}

\begin{document}

\title{Fractonic superfluids (II): condensing subdimensional particles}
\author{Shuai A. Chen}
\affiliation{Institute for Advanced Study, Tsinghua University, Beijing,
100084, China}

\author{Jian-Keng Yuan}
\affiliation{School of Physics and State Key Laboratory of Optoelectronic Materials and
Technologies, Sun Yat-sen University, Guangzhou, 510275,
China}

\author{Peng Ye}
\email{yepeng5@mail.sysu.edu.cn}
\affiliation{School of Physics and State Key Laboratory of Optoelectronic Materials and
Technologies, Sun Yat-sen University, Guangzhou, 510275,
China}
 
\begin{abstract}
As a series of work about ``fractonic superfluids'',   in this paper, we develop an exotic fractonic  superfluid  phase in $d$-dimensional space where subdimensional particles---their mobility is \emph{partially} restricted---are condensed. The off-diagonal long range order (ODLRO) is investigated. To demonstrate, we consider ``lineons''---a subdimensional particle whose mobility is free only in certain one-dimensional directions. We start with a $d$-component microscopic Hamiltonian model. The model  respects a higher-rank   symmetry such that  both particle numbers of each component and angular charge moments are conserved quantities. 
By performing the Hartree-Fock-Bogoliubov approximation, we derive  a set of Gross-Pitaevskii equations and  a Bogoliubov-de Gennes (BdG)  Hamiltonian, which leads to a description of both condensed components and unification of gapless phonons and  gapped rotons. With the coherent-path-integral representation, we also derive the long-wavelength effective field theory of  gapless Goldstone modes and analyze quantum fluctuations around classical ground states. The Euler-Lagrange equations and  Noether charges/currents are also studied. In two spatial dimensions and higher, 
such an ODLRO stays stable against quantum fluctuations. Finally, we study vortex configurations.
The higher-rank symmetry enforces a hierarchy of thermal vortex excitations 
whose structures are dominated by two guiding statements.
Specially, we construct two types of vortex excitations, 
the conventional and dipole vortices. The latter carries a charge with dimension as a momentum. 
The two statements can be more generally applicable. Further perspectives are discussed. 

\end{abstract}

\maketitle
%\tableofcontents

\section{Introduction}
As exotic states of matter, fracton topological order  can be characterized by   noise-immune ground state degeneracy that unconventionally  depends on the system size on a non-trivial compact manifold \cite{Chamon05, Haah2011, Vijay2015, Vijay2016}. Recently, fracton topological order or \emph{fracton physics} in a more general sense has been intensively investigated, see, e.g., Refs.~\cite%
{Vijay2015,Vijay2016,Prem2017,Chamon05,Vijay2015,Shirley2019,Ma2017,Haah2011,Bulmash2019,Prem2019,Bulmash2018,Tian2018,You2018,Ma2018,Slagle2017,Halasz2017,Tian2019,Shirley2018a,Slagle2019a,Shirley2018,Prem2017,Prem2018,Pai2019,Pai2019a,Sala2019,Kumar2018,Pretko2018,Pretko2017,ye19a,Ma2018,Pretko2017a,Radzihovsky2019,Dua2019,PhysRevLett.122.076403,haahthesis,PhysRevX.9.031035,2019arXiv190411530Y,2019arXiv190408424S,2019arXiv190404815K, 2019arXiv190913879W, 2019arXiv191213485W,pretko18string,pretko18localization,PhysRevB.100.125150,PhysRevB.99.245135,PhysRevB.97.144106,PhysRevB.99.155118,MaHigherRankDQC, 2019arXiv191101804W, 2020PhRvR2b3267Y, 2020arXiv200803852S, 2020arXiv200707894W, 2020arXiv200704904G, 2020arXiv200512317N, 2020arXiv200414393P, 2020arXiv200407251W, 2020arXiv200406115S, 2020arXiv200404181S,2020arXiv200400015S, 2020PhRvR2c3124G, 2020arXiv200212932W, 2020arXiv200212026S, 2020arXiv200205166A, 2020arXiv200202433W, 2020arXiv200105937P}.  A recent review can be found in Refs.~\cite{Nandkishore2019, 2020Fracton}. Topological excitations of fracton topological order include fractons, subdimensional particles \cite{Chamon05, Haah2011, Vijay2015, Vijay2016,Pretko2017} and more complicated spatially extended excitations \cite{ye19a}. One of remarkable features of these excitations is  topological restriction on mobility: Their geometrical locations cannot be freely changed by any local operators. More concretely, mobility of fractons is completely frozen while subdimensional particles are still allowed to move but within a certain cluster of lower-dimensional subspace.  As two examples of subdimensional particles in the $X$-cube lattice model \cite{Chamon05, Haah2011, Vijay2015, Vijay2016}, lineons and planeons can move along certain one-dimensional directions and   two-dimensional parallel planes, respectively.  
 Instead of the interpretation as ``topological excitations'', one can also regard all these strange particles, i.e., fractons, lineons, and planeons as original bosons, which leads to unconventional many-body physics.   In this context, the restriction on mobility is ascribed to the implementation of so-called ``higher-rank symmetry''. The latter   guarantees a set of higher moments are conserved \cite{Pretko2018, PhysRevX.9.031035, Seiberg2019arXiv1909}.

 In a previous work \cite{2020PhRvR2b3267Y} of many-body physics of fractons, the authors of the present work proposed a \emph{fractonic superfluid phase} formed by non-relativistic bosonic fractons in $d$ spatial dimensions ($d$D). The phase can be regarded as a result of spontaneous breakdown of  higher-rank symmetry. Due to a higher-rank symmetry,  both  total dipole moments and the total particle number (charge) are conserved. The microscopic Hamiltonian that respects such a symmetry must be non-Gaussian, which naturally  forbids a single fracton from freely propagating.  Starting with the first-order time derivative just like a conventional superfluid phase,  we add  
a usual Mexican-hat potential for fractons. When the  chemical potential is turned from a negative to positive value, the system 
undergoes a quantum phase transition from the normal state to the superfluid phase. The latter is manifested by occupation 
of  a macroscopic number of fractons on the same quantum state, which leads to the formation of an off-diagonal long range order (ODLRO) \cite{YangODLRO}.  As a direct consequence of non-Gaussianality, the corresponding Euler-Lagrange equation is highly non-linear, from which one can extract 
 time-dependent Gross-Pitaevskii   equation that governs hydrodynamical behaviors. Furthermore, by taking quantum phase fluctuations 
 into consideration,
   we find that   ODLRO keeps stable in spatial dimensions $d>2$. In $1$D, the correlation function of the superfluid order parameter
 exponentially decays at long distances. In  $2$D,  it decays in a power-law pattern at zero temperature.

 %%%%%%%%%%%%%%%%%%%%
\begin{table*}[tbp] \centering%
%EndExpansion
\caption{Comparison between a conventional superfluid phase (denoted as $d\mathsf{SF}^d$), fractonic superfluid phase (denoted as $d\mathsf{SF}^0$; see Ref.~\cite{2020PhRvR2b3267Y}) via  condensing fractons, and  
a fractonic superfluid phase (denoted as $d\mathsf{SF}^1$) via condensing lineons. In these three types of superfluids, the condensed particles are, respectively, usual bosons of full  mobility, fractons without any mobility, and lineons with partial mobility. Vortex excitations in $2$D form a hierarchy where $\ell,\ell_1, \ell_2$ denote winding numbers and $p,p_1, p_2$ are quantized as  momenta  with $\varphi(\mathbf x) $ being the relative angle of site $\mathbf{ x}$ to the vortex core. 
} 
\label{Tab}%

\begin{tabular}{cccc}

\hline

\hline

& $d\mathsf{SF}^{d}$ & $d\mathsf{SF}^{0}$ & $d\mathsf{SF}^1$ \\ 
\hline
Conserved quantities & Total charge & Total charge, total dipole moment & 
Total charges, angular charge moments \\ 
Order parameter & $\sqrt{\rho _{0}}e^{i\theta _{0}}$ & $\sqrt{\rho _{0}}%
e^{i\left( \theta _{0}+\sum_{a}\beta _{a}x^{a}\right) }$ & $\sqrt{\rho _{0}}%
e^{i\left( \theta _{a}+\sum_{b}\beta _{ab}x^{a}\right) }\left( \beta
_{ab}=-\beta _{ba}\right) $ \\ 
Plane-wave dispersion & Dispersive & Dispersionless & Partially dispersive
\\ 
Ground state & $e^{\int \mathrm d^{d}x\sqrt{\rho _{0}}e^{i\theta _{0}}\hat{\Phi}%
^{\dag }( \mathbf{x}) }|0\rangle $ & $e^{\int\mathrm d^{d}x\sqrt{\rho
_{0}}e^{i\left( \theta _{0}+\sum_{a}\beta _{a}x^{a}\right) }\hat{\Phi}%
^{\dag }\left( \mathbf{x}\right) }|0\rangle $ & $\prod_{a}e^{\int
\mathrm d^{d}x\sqrt{\rho _{0}}e^{i\left( \theta _{a}+\sum_{b}\beta
_{ab}x^{a}\right) }\hat{\Phi}_{a}^{\dag }( \mathbf{x}) }|0\rangle 
$ \\ 
Specific capacity heat & $c_{\upsilon }\varpropto T^{d}$ & $c_{\upsilon
}\varpropto T^{\frac{d}{2}}$ & $c_{\upsilon }\varpropto T^{d}$ \\ 
Number of Goldstone modes & $1$ & $1$ & $d$ \\
Dispersion of  Goldstone modes & $\omega \varpropto \left\vert \mathbf{k}\right\vert $ & $%
\omega \varpropto \left\vert \mathbf{k}\right\vert ^{2}$ & $\omega
\varpropto \left\vert \mathbf{k}\right\vert $ \\ 
Stable dimension at $T=0$ & $d>1$ & $d>2$ & $d>1$ \\ 
Vortex structure in $d=2$ & $\ell\varphi \left( \mathbf{x}\right) $ & $\ell\varphi
\left( \mathbf{x}\right) ,\left( p_{1}x+p_{2}y\right) \varphi \left( \mathbf{x}%
\right) $ & $%
\begin{array}{c}
\ell_{1}\varphi \left( \mathbf{x}\right) ,px^{2}\varphi \left( \mathbf{x}\right)  \\ 
\ell_{2}\varphi \left( \mathbf{x}\right) ,px^{1}\varphi \left( \mathbf{x}\right) 
\end{array}%
$ \\ \hline 

\hline

\end{tabular}%

%TCIMACRO{\TeXButton{E}{\end{table}}}%
%BeginExpansion
\end{table*}%
%EndExpansion

As mentioned above, fractons are just one of many strange particles proposed in fracton topological order. We expect that subdimensional particles can form even more   exotic phases of matter as their mobility is partially rather than completely restricted.  Along this line, in this work, we  consider more general variants of fractonic superfluids   where, instead of completely immobile fractons, subdimensional  particles  meet Mexican hats and thus form a superfluid. For convenience, we introduce a notation $d\mathsf{SF}^{n}$ (``$\mathsf{SF}$'' stands for ``superfluid'') which represents a superfluid phase in $d$ spatial dimensions via condensing subdimensional particles of dimension-$n$ ($0\leq n\leq d$). For example, $d\mathsf{SF}^{d}$ denotes a conventional superfluid where the bosons are free to move in the whole space and the fractonic superfluid phase in a many-fracton model \cite{2020PhRvR2b3267Y} is symbolized as $d\mathsf{SF}^{0}$.  

More specifically, in this work, we take lineons as an example, which leads to a fractonic superfluid phase denoted by $d\mathsf{SF}^1$.   The corresponding microscopic second-quantized Hamiltonian model $\mathcal{H}$ contains $d$ components  of bosonic lineons. 
The Hamiltonian has strongly anistropic quadratic Gaussian terms such that mobility restriction of lineons is correctly encoded. 
Meanwhile, both  angular charge moments and particle numbers of each component are conserved due to the presence of quartic terms  that respect a higher-rank symmetry.  
The candidate Hamiltonian is referred to in  Ref.~\cite{Pretko2018} with the second time-derivative terms. Distinguishably,  we  set about the first-time derivative and  apply a Mexican-hat potential to each component. In the coherent-path-integral representation, the Hamiltonian density  $\mathcal H$ is sent to the Lagrangian density $\mathcal{L}=\sum_{a=1}^d i\phi_{a }^*\partial_t\phi_a^{}-\mathcal H$ after a Wick rotation. Due to the first-order time-derivative, we are legitimate to interpret $\phi_a^*\phi_a^{}$ as the particle density of the $a^{\text{th}}$  component, which is common   in   condensed matter   and cold-atom systems.

When the chemical potential is turned to a positive value from a negative one, a quantum phase transition occurs from the normal state to the fractonic superfluid phase $d\mathsf{SF}^1 $. The normal state has a unique ground state with  vanishing momentum. Instead, in $d\mathsf{SF}^1 $, ODLRO 
the ground states are macroscopically degenerate in the classical level and  their configurations appear to be a plane-wave with a finite momentum and a finite density distribution which establishes a ODLRO. Upon the   {Hartree-Fock-Bogoliubov}   mean-field approximation is applied,  the boson fields are further split into two components---the normal and the condensed components,  towards which we derive  a set of non-linear Bogoliubov-de Gennes   equations and Gross-Pitaevskii   Hamiltonian respectively. To deal with gapless Goldstone modes  and quantum phase fluctuations (i.e., ODLRO stability at infrared limit), we turn to the framework of  an effective field theory. In contrast to the non-Gaussian system in Ref.~\cite{2020PhRvR2b3267Y}, existence of spatially anisotropic Gaussianality ensures that a superfluid phase or ODLRO can survive against quantum fluctuations in spatial dimensions $d>1$ at zero temperature and thus this model appears more tractable experimentally. 
For vortex configurations in $2\mathsf{SF}^1$, we point out and apply two statements for the purpose of  constructing  vortex configurations. 
The first statement dominates the multi-valued part to meet the single-valueness of vortex fields and the second controls the smooth part to satisfy a relation between representations of a higher-rank group  and a  particle number conservation symmetry group.
The two statements lead to two types of vortices: the conventional  vortex and dipole vortex. The latter carries a dipole charge that is quantized like a momentum. It can be detected by a vorticity from recombination between Noether currents. In fact, the two statements can be applied to 
point vortex excitations with a general higher-rank symmetry. In Table.~\ref{Tab}, we compare different properties of 
a conventional superfluid phase $2\mathsf{SF}^2$, fractonic superfluid via condensing fractons $2\mathsf{SF}^0$ in Ref.~\cite{2020PhRvR2b3267Y} and  fractonic superfluid by condensing lineons $2\mathsf{SF}^{1}$ in this work.

The remaining part of this paper is organized as follows. Sec.~\ref{Sec::Intro} provides   a microscopic multi-component model and Hartree-Fork-Bogoliubov treatment. An effective field theoretical analysis is performed in Sec.~\ref{section_eft111}.  In Sec.~\ref{Sec:Topovortex}, exotic superfluid vortices are studied. This work is concluded in Sec.~\ref{Sec::concl}.

%%%%%%%%%%%%%%%%%%%%%%%%%
\section{Microscopic system and Mean-field theory}
\label{Sec::Intro}
In this section,  we start with a microscopic model of lineons, which is formulated in the second quantization language with conserved angular charge moments. This conservation is vital  to  the  mobility restriction of    lineons. Under the circumstance of condensing lineons, we apply the Hartree--Fock--Bogoliubov (HFB) mean-field theory to derive the Gross--Pitaevskii (GP) equations and
Bogoliubov--de Gennes (BdG) Hamiltonian  \cite{fetter1971quantum}.
 The former govern the order parameter of the superfluid phase and the latter unifies both gapless phonon and gapped roton modes.

\subsection{A model Hamiltonian}\label{section_subsub}

\begin{figure}[t]
\centering 
\includegraphics[scale=0.05]{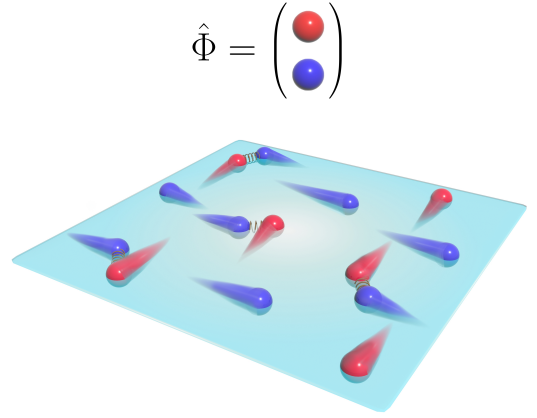}
\caption{Illustration of the interacting system in two dimensions. The red and blue balls respectively represent two components $a=1,2$, which move along distinct orthogonal directions. The spring between two balls 
represents the interaction due to the $K$-term in Eq.~(\ref{Ham}).
}
\label{Fig2D}
\end{figure}

In  condensed matter systems, strong anisotropy can constrain particle's
propagation.  For example, divergence of effective mass localizes particle spatially and
a strong electric field allows charged particles to move  exclusively along the direction of electric field.\
We have investigated one microscopic realization of fractons with fully restricted motion in Ref.~\cite{2020PhRvR2b3267Y}. As a series of works, here we focus on $d$-component fields $\hat{\Phi}=(\hat \Phi _{1},\cdots, \hat\Phi
_{d}) $ in $d$ spatial dimensions in the Hamiltonian $H=\int \mathrm d^{d}%
\mathbf{x}\mathcal{H}$ where Hamiltonian density $\mathcal{H}$ reads
\begin{align}
\mathcal{H} =&\sum_{a=1}^{d}\partial _{a}\hat{\Phi}_{a}^{\dag }\partial _{a}%
\hat{\Phi}_{a}  \notag \\
&+\sum_{a\not=b}^{d}\frac{1}{2}K _{ab}(\hat{\Phi}_{a}^{\dag }\partial
_{a}\hat{\Phi}_{b}^{\dag }+\hat{\Phi}_{b}^{\dag }\partial _{b}\hat{\Phi}%
_{a}^{\dag })\left( \hat \Phi _{a}\partial _{a}\hat \Phi _{b}+\hat \Phi _{b}\partial
_{b}\hat \Phi _{a}\right)  \notag \\
&+V(\hat{\Phi}^{\dag },\hat{\Phi})~.  \label{Ham}
\end{align}%
The complex fields $\hat{\Phi}_{a}^{\dag }\left( \mathbf{x}\right) $ and $%
\hat{\Phi}_{b}\left( \mathbf{x}\right) $ create and annihilate an 
$a$th-component particle respectively and satisfy the bosonic communication
relations%
\begin{equation}
[ \hat{\Phi}_{a}\left( \mathbf{x}\right) ,\hat{\Phi}_{b}^{\dag }\left(
\mathbf{y}\right) ] =\delta _{ab}\delta ^{d}\left( \mathbf{x-y}\right)\,,
\label{commrelation}
\end{equation}%
where $\mathbf{x}=(x^1,\cdots, x^d)$ is the spatial coordinate. At the quadratic level, each
component can only propagate in one certain spatial directions and we set the mass before the quadratic terms to be a unit.  For convenience, one can
set diagonal terms $K _{aa}=0$ for $a=1,\cdots, d$ since diagonal terms $K_{aa}$ is  absent  in Eq.~\eqref{Ham}.
For simplicity, we can take the term $V(\hat{\Phi}^\dag,\hat{\Phi})$ to be the Mexican-hat potential with component-independent chemical potential $\mu$ and interaction coupling constant $g>0$ ,
\begin{equation}
V(\hat{\Phi}^{\dag },\hat{\Phi})=\sum_{a=1}^{d}-\mu \hat{\Phi}_{a}^{\dag
}\hat{\Phi}_{a}+\frac{g}{2}\hat{\Phi}_{a}^{\dag }\hat{\Phi}_{a}^{\dag }%
\hat{\Phi}_{a}\hat{\Phi}_{a} \label{potentialV}
\end{equation}
which describes a short-range repulsive
interaction via the s-wave scattering. In the following, \emph{no Einstein
summation rule is assumed.} 
The Hamiltonian in Eq.~(\ref{Ham}) conserves  both particle numbers of each components $Q_a\equiv \int \mathrm d^dx\hat{\rho}_a$  and angular charge moments $Q_{ab}=\int \mathrm d x^d (\hat \rho_a x^b-\hat \rho_b x^a)$ with $\hat{\rho}_a=\hat{\Phi}_a^\dag \hat{\Phi}_a^{}$ being number operator of $a$th particles. Accordingly, the symmetry group is composed of transformations $\hat\Phi_a\rightarrow e^{i\lambda_a}\hat\Phi_a$ for each component  and 
\begin{equation}
( \hat{\Phi}_{a},\hat{\Phi}_{b}) \rightarrow ( \hat{\Phi}%
_{a}e^{i\lambda _{ab}x^{b}},\hat{\Phi}_{b}e^{-i\lambda _{ab}x^{a}})
\label{momentsymm}
\end{equation} 
for each pair of indices with $\lambda_a, \lambda_{ab}\in \mathbb{R}$. The parameters $\lambda_{ab}$ are anti-symmetric $\lambda_{ab}=-\lambda_{ba}$,  thus inducing $\frac{d(d-1)}{2}$ independent conserved angular charge moments $Q_{ab}$. 
The transformations in Eq.~(\ref{momentsymm}) involve local coordinates $\mathbf x=(x^1,x^2,\cdots, x^d)$, and they do not form an internal symmetry. We denote the symmetry group as $\mathcal G$ which characterizes a higher-rank symmetry \cite{Seiberg2019arXiv1909} . 
In a periodic boundary condition, the parameters $\lambda_{ab}$ have the dimension of $[x]^{-1}$ and quantization of the related charges is expected to
 coincide with a momentum.
This symmetry intertwines global and internal
symmetries such that strong constraints are imposed on particles' propagations.
 In $2$D, conservation of $%
Q_{12}=\int \mathrm d^{2}x\left( \hat{\rho}_{1}x^{2}-\hat{\rho}_{2}x^{1}\right) $
requires the velocity shall be parallel to a vector $\left( \hat{\rho}_{1},%
\hat{\rho}_{2}\right) $ as a lineon. In three spatial dimensions, we have $3$ angular charge
moments $Q_{12},Q_{23},Q_{13},$ such that a particle only
propagates in the direction parallel to $\left( \hat{\rho}_{1},\hat{\rho}%
_{2},\hat{\rho}_{3}\right) $.
Generally,  fundamental particles in $d$D move with velocity parallel to $\left( \hat{\rho} _{1},\cdots
,\hat\rho _{d}\right) $.  
Fig.~\ref{Fig2D} pictorially shows the interacting system when $d=2$.
%%%

\subsection{Hartree-Fock-Bogoliubov mean-field theory: condensate and rotons\label{Sec:MeanFT}}

It is well-known that a Bose-Einstein condensate  consists of a
two-component structure: the condensate and the normal components. 
The HFB mean-field theory allows
factorization of fields $\hat{\Phi}_{a}$ in terms of an appropriate
orthonormal single-particle basis,
\begin{equation}
\hat{\Phi}_{a}( \mathbf{x}) =\phi _{a0}\left( \mathbf{x}\right)
\hat{c}_{a0}+\sum_{i\not=0}\phi _{ai}\left( \mathbf{x}\right) \hat{c}%
_{ai}\equiv \phi _{a0}\left( \mathbf{x}\right) \hat{c}_{a0}+\hat{\psi}%
_{a}\left( \mathbf{x}\right)\,,  \label{phiExpansion}
\end{equation}%
with $\hat{\psi}_{a}\left( \mathbf{x}\right) \equiv \sum_{i\not =0}\phi
_{ai}\left( \mathbf{x}\right) \hat{c}_{ai}$ where the operator $\hat{c}_{ai}$
satisfies the bosonic commutation relations $[\hat{c}_{ai},\hat{c}%
_{bj}^{\dag }]=\delta _{ab}\delta _{ij}$.
The wavefunctions $\phi _{a0}\left( \mathbf{x}\right) $ signify the condensate component. 
 The non-condensate fields $\hat{c}_{ai}\left(
i\geq 1\right) $ constitute a branch of gapped quasiparticle excitations
and are orthogonal with the condensate component $\phi _{a0}\left( \mathbf{x%
}\right) ,$
\begin{equation}
\int \mathrm d^{d}\mathbf{x}\phi _{a0}( \mathbf{x}) \phi _{bi}^{\ast
}( \mathbf{x}) =0\text{ for }i\geq 1 ~. \label{orthogonal}
\end{equation}%
Here the subscript index `$i$' can be taken as a spectral index of the BdG Hamiltonian [see Eq.~\eqref{BdGeq} ], and thus $\phi_{ai}$ can denote wave functions of low-energy excitations. 
We take the normal component as  a perturbation to the condensate. 
 Substituting Eq.~(\ref{phiExpansion}) into
Hamiltonian in Eq.~(\ref{Ham}) leads to a partition into terms with
different numbers of field operators $\hat{\psi}_{a}( \mathbf{x})
$. The zeroth-order term is given by
\begin{equation}
H_{0}=\sum_{a=1}^{d}h_{1,a}\hat{N}_{a0}+\sum_{a,b}h_{2,ab}(\hat{N}_{a0}\hat{N%
}_{b0}-\delta _{ab}\hat{N}_{a0})~,  \label{H0}
\end{equation}%
where $\hat{N}_{a0}=\hat{c}_{a0}^{\dag }\hat{c}_{a0}^{{}}$ is the number
operator for the condensate and $h_{1}\left( h_{2}\right) $ is the
expectation value of the single-particle (two-particle) part of the
Hamiltonian
\begin{align}
h_{1,a} =&\int \mathrm d^{d}\mathbf{x}\left\vert \partial _{a}\phi _{a0}\right\vert
^{2}-\mu \left\vert \phi _{a0}\right\vert ^{2} ~,\\
h_{2,ab}  =&\int  \mathrm d^{d}\mathbf{x}\frac{1}{2}K _{ab}|\phi _{a0}\partial
_{a}\phi _{b0}+\phi _{b0}\partial _{b}\phi _{a0}|^{2}\nonumber\\
&+\frac{g\delta _{ab}}{2}%
\left\vert \phi _{a0}\phi _{b0}\right\vert ^{2}.
\end{align}%
Under the particle-number representation $|\left\{ N_{a0}\right\} \rangle $
with definite condensate particle number $\hat{N}_{b0}|\left\{
N_{a0}\right\} \rangle =N_{b0}|\left\{ N_{a0}\right\} \rangle $ for $%
b=1,\cdots ,d$, the ground state energy $%
E_{0}\left( \left\{ N_{a}\right\} \right) $ of $H_{0}$ in Eq.~(\ref{H0}) only depends on the
condensate,
\begin{align}
E_{0}\left( \left\{ N_{a}\right\} \right) &=\left\langle \left\{
N_{a0}\right\} |H_{0}|\left\{ N_{a0}\right\} \right\rangle  \notag \\
&=\sum_{a}h_{1,a}N_{a0}+\sum_{a,b}h_{2,ab}(N_{a0}N_{b0}-\delta _{ab}N_{a0})~.
\end{align}%
The next order $H_{1}$ has linear dependence on $\hat{\psi}_{a}\left(
\mathbf{x}\right) $%
\begin{equation}
H_{1}=\int  \mathrm d^{d}\mathbf{x}\sum_{a=1}^{d}\hat{c}_{a0}^{\dag }\phi _{a0}^{\ast
}\mathcal{\hat{H}}_{a}^{\dag }\hat{\psi}_{a}+\hat{\psi}_{a}^{\dag }\mathcal{%
\hat{H}}_{a}\phi _{a0}\hat{c}_{a0}~,
\end{equation}%
where%
\begin{equation}
\mathcal{\hat{H}}_{a}=-\partial _{a}^{2}-\mu +g\left\vert \phi
_{a0}\right\vert ^{2}\hat{N}_{a0}+\sum_{a,b}\frac{1}{2}K _{ab}\hat{N}%
_{b0}\mathcal{\hat{H}}_{ab}  \label{Ha}
\end{equation}%
with
\begin{align*}
\mathcal{\hat{H}}_{ab}=&(\partial _{a}\phi _{b0}^{\ast })(\partial _{a}\phi
_{b0})+(\partial _{a}\phi _{b0}^{\ast })\phi _{b0}\partial _{b}-(\partial
_{b}\phi _{b0}^{\ast })(\partial _{a}\phi _{b0}) \notag \\
&-\phi _{b0}^{\ast }(\partial
_{a}\phi _{b0})\partial _{b} -\phi _{b0}^{\ast }(\partial _{a}\partial _{b}\phi _{b0})-(\partial _{b}\phi
_{b0}^{\ast })\phi _{b0}\partial _{b} \notag \\
& -\phi _{b0}^{\ast }(\partial _{b}\phi
_{b0})\partial _{b}-\phi _{b0}^{\ast }\phi _{b0}\partial _{b}^{2}~.
\end{align*}%
Under a basis $|\left\{ N_{a0}\right\} \rangle $, taking the limit of $%
N_{a0}\gg 1$, one can recognize$\ \hat{N}_{a0}\hat{c}_{a0}|\left\{
N_{a}\right\} \rangle =\hat{c}_{a0}\hat{N}_{a0}|\left\{ N_{a}\right\}
\rangle $. If $\phi _{a0}\left( \mathbf{x}\right) $ are chosen to be eigenstates of the operator $\mathcal{\hat{H}}_{a}$ in Eq.~\eqref{Ha},
\begin{equation}
\mathcal{\hat{H}}_{a}\left( \mathbf{x}\right) \phi _{a0}\left( \mathbf{x}%
\right) =\epsilon _{a0}\left( \left\{ N_{a0}\right\} \right) \phi
_{a0}\left( \mathbf{x}\right) \,\, a=1,\cdots d~, \label{GPeq}
\end{equation}%
 then $H_{1}$ vanishes
identically due to orthogonality in Eq.~(\ref{orthogonal}) and $N_{a0}\gg 1$, where ground state energy $\epsilon _{a0}\left( \left\{ N_{a0}\right\}
\right) $ merely depends on the condensate component.
This fact establishes the validity of the expansion in Eq.~(\ref%
{phiExpansion}). The Eq.~(\ref{GPeq}) marks a set of the  GP equations %
describing the condensate components where $\mathcal{\hat{H}}_{a}$ in Eq.~(%
\ref{Ha}) behaves as a single-particle Hamiltonian. It simply directs us to
approximate the original Hamiltonian in Eq.~(\ref{Ham}) by Eq.~(\ref{H0}). 
With the translational symmetry, the GP equations have a set of very simple solutions.
If the chemical potential $\mu$ is negative, the ground state energy $E_0$ reaches its minimum 
when the condensate has vanishing density, $ N_{a0}=0$ for $a=1,\cdots d$. We obtain a normal state.
If the chemical potential is switched to a positive value, the ground state energy $E_0$ reaches maximum with a finite number of the condensate component.
In this case, the configurations of ground states can be
parametrized by real parameters $\theta _{a}$ and $\beta _{ab}$ $( \beta
_{ab}=-\beta _{ba}, a,b=1,\cdots d) $,
\begin{equation}
\phi _{a0}\left( \mathbf{x}\right) =\frac{1}{\sqrt{V}}e^{i\left( \theta
_{a}+\sum_{b=1}^{d}\beta _{ab}x^{b}\right) } ~, \label{phia0}
\end{equation}%
with $V$ as the spatial volume. 
 Remarkably, Eq.~\eqref{phia0} depends on the parameter $\beta_{ab}$ carrying the dimension of `momentum'. In other words,     
the condensate component can carry finite momentum. 
Of course, one can include the trap potential
that can break a translational symmetry, under which the GP equations may be short of
analytical solutions.
Casting the solution in Eq.~(\ref{phia0}) back to $%
H_{0}$ in Eq.~(\ref{H0}), we have the ground state energy $E_{0}\left(
\left\{ N_{a0}\right\} \right) $ ,
\begin{equation}
E_{0}\left( \left\{ N_{a0}\right\} \right) =\sum_{a=1}^d-\mu \frac{N_{a0}}{V}+%
\frac{g}{2}\left( \frac{N_{a0}}{V}\right) ^{2}~.
\end{equation}
The minimal condition of $E_{0}\left( \left\{ N_{a0}\right\} \right) $ with
a positive chemical potential $\mu $ requires $\rho _{a0}\equiv \frac{N_{a0}}{V}=\frac{\mu }{g}$%
, which indicates the ground states for a fractonic superfluid has a
macroscopically finite particle density. Hence, we obtain a superfluid phase by condensing lineons, which we dub   $d\mathsf{SF}^1$.   
Therefore, select one ground state
in Eq.~(\ref{phia0}) and we can fix the condensate particle number $N_{a0}$ 
by simply replacing both $\hat{c}_{a0}$ and $\hat{c}_{a0}^{\dag }$ operators
by c-number $\sqrt{N_{a0}}$, which indicates occurrence of ODLRO with the condensate density $\rho _{a0}$.

%%%%%%%%%%%%%%
\begin{figure}
\centering 
\includegraphics[scale=0.33]{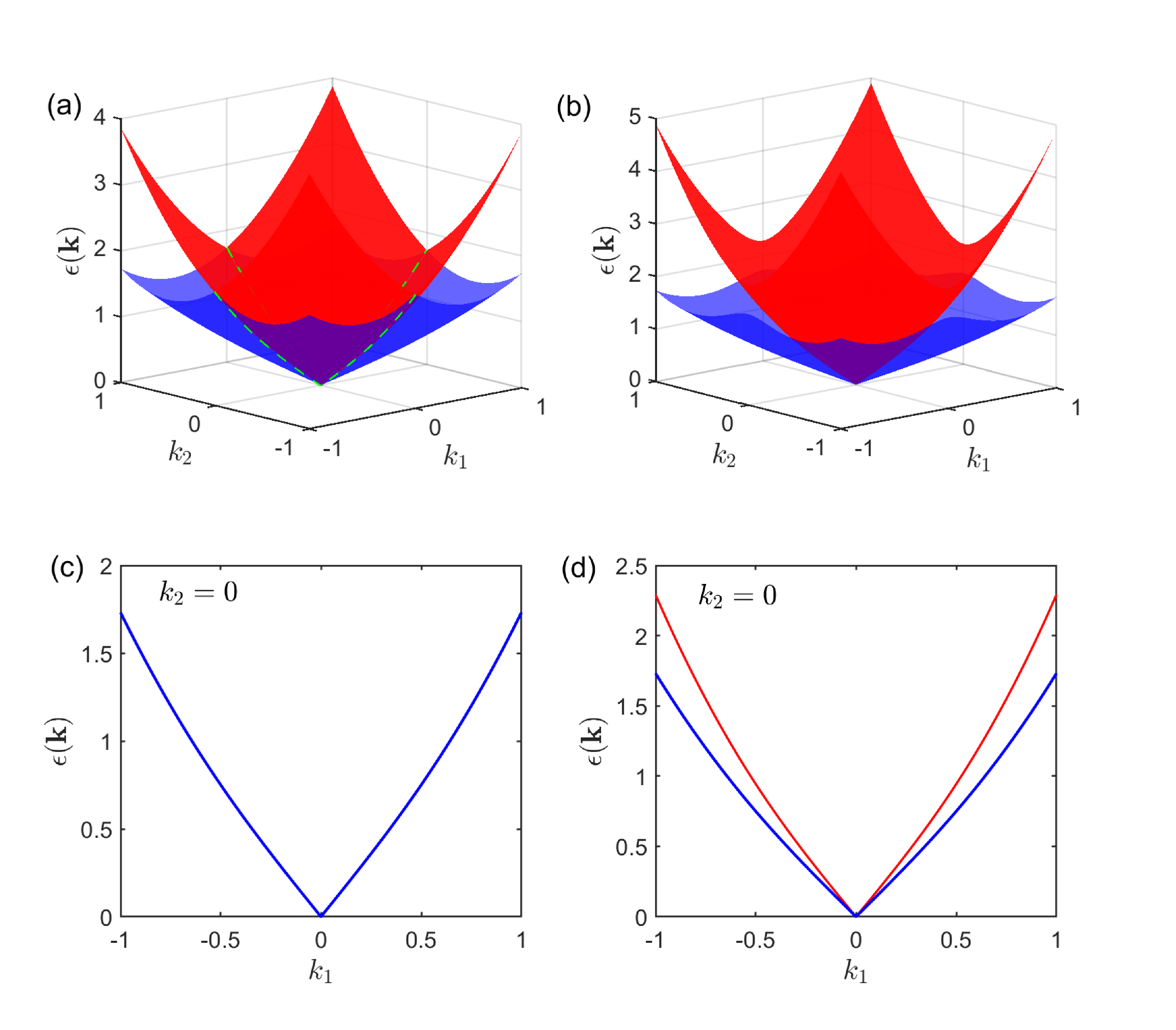}
\caption{Spectrum dispersions of quasiparticles from HFB approximation in Eq.~\eqref{disp2D} in 2D with parameters 
in (a),(c) $K=1,\rho_0=1$ and (b),(d) $K=1.5,\rho_0=1$.  In (a) the two bands are degenerate at two lines $k_1=0$ and $ k_2=0$ which are guided by green dashed lines. 
The gap between the two bands in (b) keeps finite except at the point $\mathbf k=\mathbf 0$.  
(c) and (d) show the dispersion relations at the line $k_2=0$. 
Both in (a) and (b), the dispersions are linear at small $\mathbf k$ and quadratic at large $\mathbf k$. 
 }
 \label{Fig1}
\end{figure}
%%%%%%%%

The next order $H_{2}$ goes beyond the GP equation to include the quadratic terms of $%
\psi _{a}$,
\begin{align}
H_{2} &=\sum_{a=1}^{d}\int \mathrm d^{d}\mathbf{x}\hat \psi _{a}^{\dag }( -\partial
_{a}^{2}+g\rho _{a0}) \hat\psi _{a}  \notag \\
&+\sum_{a,b}\frac{1}{2}K _{ab}\rho _{a0}\left( \partial _{a}\hat\psi
_{b}^{\dag }+\partial _{b}\hat\psi _{a}^{\dag }\right) \left( \partial _{a}\hat\psi
_{b}+\partial _{b}\hat\psi _{a}\right)  \notag \\
&+\sum_{a=1}^{d}\frac{g}{2}\rho _{a0}\left( \hat\psi _{a}^{\dag }\hat\psi
_{a}^{\dag }+\hat\psi _{a}\hat\psi _{a}\right) \,, \label{BdGeq}
\end{align}%
where we have replaced $\hat{c}_{a0}$ and $\hat{c}_{a0}^{\dag }$ with $\sqrt{N_{a0}%
}$. New quadratic terms emerge from $K _{ab}$-term which relaxes the
restrictions on dynamics. It means that the quasiparticle modes $\hat \psi _{a}$
can propagate along all spatial directions. The mass term $g\rho _{a0}\hat\psi
_{a}^{\dag }\hat\psi _{a}$ originates from the condensate component. 
$H_{2}$ in Eq.~\eqref{BdGeq} is designated as a  BdG Hamiltonian to
characterize the non-condensate quasiparticle modes.
One can diagonalize $%
H_{2}$ to obtain the canonical quaisparticle modes. For example, in $2$D, the spectrum has two branches,
\begin{equation}
\epsilon_{\pm}=\sqrt{\left[\frac{(1+K\rho_0)|\mathbf k|^2}{2} \pm\frac{ \Delta(\mathbf k)}{2}+g\rho_0\right]^2-(g\rho_0)^2 }~.
\label{disp2D}
\end{equation}
For small momentum, up to the first order, we have linear gapless dispersion relations, 
\begin{equation}
\epsilon_{\pm}(\mathbf k)=\sqrt{g\rho_0} \sqrt{(1+K\rho_0)\mathbf k^2\pm  \Delta(\mathbf k) }\quad\mathrm{small}~\mathbf k ~,
\label{BdGdispersion1}
\end{equation}
and they describe two gapless phonon excitations, 
while at the large momentum,  spectrum dispersions depend on momentum quadratically,
\begin{align}
\epsilon_{\pm}=\frac{(1+K\rho_0)|\mathbf k|^2}{2} \pm \frac{\Delta(\mathbf k)}{2}\quad\mathrm{large}~\mathbf k~,
 \label{BdGdispersion2}
\end{align}
and instead they correspond to gapped roton modes. Here   $
\Delta(\mathbf k)\equiv  \sqrt{\left( k_{1}^{2}-k_{2}^{2}\right)
^{2}\left( K \rho _{0}-1\right) ^{2}+4K ^{2}\rho
_{0}^{2}k_{1}^{2}k_{2}^{2}}$, $\rho_0=\frac{\mu}{g}$ and $K\equiv K_{12}$. 
The smooth change from linear to quartic dispersion is the key feature of the HFB approximation. 
Although two modes, phonons and rotons, are emphasized, indeed they represent different behavors at small and high momentum respectively. 
The splitting between the two dispersion relations 
in Eqs.~\eqref{BdGdispersion1} and \eqref{BdGdispersion2} is controlled by $ \Delta(\mathbf k)$. 
%The dispersion relations in Eqs.~\eqref{BdGdispersion1} and \eqref{BdGdispersion2} have a \emph{band splitting gap} (BSG) controlled by $%
%\Delta (\mathbf k)$. 
If $K\rho _{0}=1$ such that $\Delta(\mathbf k) =0$,  the two bands are degenerate at two lines $k_{1}=0$ or 
$k_{2}=0$ and along the two lines, the bands have ill-defined curvatures, which is expected to be detected by thermal Hall effect. Figure~\ref{Fig1} depicts the two dispersion relations in Eq.~\eqref{disp2D} in which the two cases  of
 $K\rho_0=1 $ or not. 

Higher-order terms couple the normal with the condensate part and describe the interaction between phonon modes, which is
beyond the  scope of this work and we leave it to future work.

\section{Effective field theory}\label{section_eft111}
The HFB mean-field method unifies  gapless phonon modes and  gapped roton modes via a BdG Hamiltonian in Eq.~(\ref{BdGeq}) in a fractonic superfluid phase $d\mathsf{SF}^1$. 
Nevertheless, gapless mode excitations can destroy BEC or ODLRO.
 In this section, we deal with gapless modes of $d\mathsf{SF}^1$ in
the framework of a continuous field theory and discuss stability of $d\mathsf{SF}^1$  against quantum fluctuations. 

\subsection{Euler-Lagrange equation and Noether charge/current}

For the coherence and completeness of the present section, we re-derive some
quantities from the field-theoretical perspective.

We perform a coherent-state path integral quantization \cite{Altland_simons_2010} to get the Lagrangian
density $\mathcal{L}$,
\begin{equation}
\mathcal{L}=\sum_{a=1}^{d}i\phi _{a}^{\ast }\partial _{t}\phi _{a}-\mathcal{H%
}\left( \phi \right)~,  \label{lagrangianden}
\end{equation}%
where $\phi _{a}( \mathbf{x},t) $ is an eigenvalue of $\hat{\Phi}%
_{a}( \mathbf{x}) $ on a coherent state $\hat{\Phi}_{a}(
\mathbf{x}) |\phi _{a}( \mathbf{x},t ) \rangle =\phi
_{a}( \mathbf{x},t) |\phi _{a}( \mathbf{x},t) \rangle $.
The first order derivative in Eq.~(\ref{lagrangianden}) in nature is
determined by the commutation relation in Eq.~(\ref{commrelation}) which can
be confirmed through the canonical quantization.
For convenience, we apply the Wick's rotation to an imaginary time at zero temperature $T=0$.

Next, we can derive the Euler-Lagrange equations as well as the Noether
currents associated with two types of conserved quantities. The
Euler-Lagrange equations can be derived from the formula $\frac{\delta
\mathcal{L}}{\delta \phi _{a}^{\ast }}=0$, explicitly,
$
i\partial _{t}\phi _{a}=\hat{H}_{a}\phi _{a}~( a=1,\cdots , d)
$, where $\hat{H}_{a}$ has the same form as $\mathcal{H}_{a}$ in Eq.~(\ref{Ha}%
),
\begin{equation}
\hat{H}_{a}=-\partial _{a}^{2}-\mu +g\left\vert \phi _{a}\right\vert ^{2}+%
\frac{1}{2}\sum_{b}\mathcal{H}_{ab}~.
\end{equation}%
Here $\mathcal{H}_{ab}$ comes from the $K _{ab}$-term,
\begin{align}
 \mathcal{H}_{ab}  = & K _{ab}\partial _{a}\phi _{b}^{\ast }\partial _{a}\phi _{b}+K
_{ab}\partial _{a}\phi _{b}^{\ast }\phi _{b}\partial _{b}-K
_{ab}\partial _{b}\phi _{b}^{\ast }\partial _{a}\phi _{b}    \notag \\
 &-K _{ab}\phi
_{b}^{\ast }\partial _{a}\phi _{b}\partial _{b} -K _{ab}\phi _{b}^{\ast }\partial _{a}\partial _{b}\phi _{b}-K
_{ab}\partial _{b}\phi _{b}^{\ast }\phi _{b}\partial _{b} \notag \\
&-K _{ab}\phi
_{b}^{\ast }\partial _{b}\phi _{b}\partial _{b}-K _{ab}\phi _{b}^{\ast
}\phi _{b}\partial _{b}^{2}\,.
\end{align}%
The Euler-Lagrange equations just recover the GP equations.
Here $\phi_a$ plays the same role of representing the condensate component as $\phi_{a0}$ in Eq.~\eqref{GPeq}.

The Hamiltonian in Eq.~(\ref{Ham}) stays invariant under transformation in
Eq.~(\ref{momentsymm}) as well as the particle number conservation symmetry. For the 
infinitesimal change $\delta \phi _{a}=i\alpha _{a}\phi _{a}$, we have the
related Noether charge $Q_{a}$ with charge density $\rho_{a}$ and currents $%
J_{i}^{a}$ that read,
\begin{align}
Q^{a}  =&\int \mathrm d^{d}\mathbf{x}\phi _{a}^{\ast }\phi _{a}\equiv \int \mathrm d^{d}%
\mathbf{x}\rho_{a}  \label{u(1)charge} \\
J_{i}^{a}  =&  iK _{ai}\rho _{a}\left(\phi _{i}\partial _{a}\phi _{i}^{\ast }- \phi _{i}^{\ast }\partial _{a}\phi
_{i}\right) \nonumber\\
&+iK _{ai}\rho
_{i}\left(\phi _{a}\partial
_{i}\phi _{a}^{\ast }- \phi _{a}^{\ast }\partial _{i}\phi _{a}\right)  \notag \\
&+i\left(\phi _{i}\partial
_{i}\phi _{i}^{\ast }- \phi _{i}^{\ast }\partial _{i}\phi _{i}\right) \delta _{ai}  \label{u(1)current}
\end{align}%
which satisfies the continuity equations $\partial _{t}\rho
^{a}+\sum_{i}\partial _{i}J_{i}^{a}=0$. Here in Eq.~\eqref{u(1)charge}, coincidence between $\phi_a^*\phi_a^{}$ and particle
density $\rho_a$ arises from the first-order time derivative in Hamiltonian in Eq.~\eqref{Ham}.
 For the transformation $\delta \phi
_{a}=ix_{b}\phi _{a}$ and $\delta \phi _{b}=-ix_{a}\phi _{b}$ corresponding
to Eq.~(\ref{momentsymm}),  we have conserved angular moments $Q_{ab}$ (with density $\rho_{ab}$) and currents $D_i^{ab}$,
\begin{align}
Q_{ab} &=\int \mathrm d^{d}x\left( \rho _{a}x^{b}-\rho_{b}x^{a}\right) \equiv \int \mathrm d^{d} x \rho_{ab} \\
D_{i}^{ab} &=x^{b}J_{i}^{a}-x^{a}J_{i}^{b}  \label{dipolecurrent}
\end{align}%
with $\rho_{a}$ and $J_{i}^{a}$ as $U(1) $ charge and current
in Eqs.~(\ref{u(1)charge}) and (\ref{u(1)current}).  The continuity equation
$
\partial _{t}\rho^{ab}+\sum_{i=1}^{d}\partial _{i}D_{i}^{ab}=0$
%\label{conserlawdipole}
%\end{equation}%
is automatically satisfied as long as the currents $J_{b}^{a}$
obey the relations $J_{b}^{a}=J_{a}^{b}$.

\subsection{Goldstone modes and quantum fluctuations}
The HFB mean-field theory in Sec.~\ref{Sec:MeanFT} starts
with one of the classical field configurations $\phi _{a}^{\textrm{cl}%
}=\sqrt{\rho _{0}}e^{(i\theta _{a}+i\sum_{b=1}^{d}\beta _{ab}x^{b})}$ which
can be formulated in the second quantization language as%
\begin{align}
\!\!\!\!|\text{GS}\rangle _{\theta _{a}}^{\beta _{ab}}=\prod\limits_{a=1}^{d}\exp [%
\sqrt{\rho _{0}}e^{i\left( \theta _{a}+\sum_{b=1}^{d}\beta
_{ab}x^{b}\right) }\hat{\Phi}_{a}^{\dag }\left( \mathbf{x}\right) ]|0\rangle~,
\label{gs}
\end{align}%
where $\hat \Phi _{a}^{\dag }( \mathbf{x}) $ $\left( a=1,\cdots
,d\right) $ creates an $a$th-component lineon with restricted motion. The
salience of Eq.~(\ref{gs}) features a finite expectation value of operator $%
\hat \Phi_{a}( \mathbf{x}) $
\begin{equation}
\langle \mathrm{GS}|
\hat \Phi _{a}( \mathbf{x}) |\mathrm{GS}\rangle
_{\theta _{a}}^{\beta _{ab}}=\sqrt{\rho _{0}}\exp (i\theta
_{a}+i\sum_{b}\beta _{ab}x^{b})~,
\end{equation}%
thus marking an ODLRO and we obtain a fractonic superfluid phase $d\mathsf{SF}^1$.
The expectation value oscillates as a plane-wave with fixed momentum $\mathbf{k}_{a}=\left(
\beta _{a1},\beta _{a2},\cdots ,\beta _{ad}\right) $ for the $a$%
-component particle. In this sense, we can rewrite $|\mathrm{GS}\rangle _{\theta
_{a}}^{\beta _{ab}}=\prod\limits_{a=1}^{d}\exp \left[ \sqrt{\rho _{0}}%
e^{i\theta _{a}}\hat \Phi _{a}^{\dag }( \mathbf{k}_{a}) \right]
|0\rangle $ with $\hat\Phi _{a}^{\dag }\left( \mathbf{k}_{a}\right) $ being the
Fourier transformation of $\hat \Phi _{a}^{\dag }( \mathbf{x}) $.
These features arise from restricted mobility of condensed particles. 
As a side note, the ground state in Eq.~\eqref{gs}, which carries finite momentum, looks like a Fulde–Ferrell–Larkin–Ovchinnikov  (FFLO) state \cite{PhysRev135A550, larkin:1964zz}. However,  the general FFLO states are formed  by Cooper pairs of \textit{fermions}
condensing at finite momentum. In contrast, the superfluid phase in Eq.~\eqref{gs} are directly formed by condensation of 
multi-component \textit{bosonic} lineons. The degenerate ground state subspace is generated by group elements of higher-rank symmetry group $\mathcal G$ (see Sec.~\ref{section_subsub}).

After condensation, the Noether currents in Eqs.~(\ref{u(1)current}) and (%
\ref{dipolecurrent}) reduce to simpler forms by expanding the field $\phi
_{a}=\phi _{a}^{\text{cl}}e^{i\theta _{a}}$ where $\phi _{a}^{\text{cl}}$
denotes classical configurations and $\theta _{a}$ are the quantum phase
fluctuations,
\begin{align}
\rho_{a} =\rho _{0} \,,
J_{i}^{a} =2K \rho _{0}^{2}\left( \partial _{i}\theta _{a}+\partial
_{a}\theta _{i}\right) +2\rho _{0}\partial _{i}\theta _{a}\delta _{ai}\text{ }
\label{condenU(1)}
\end{align}
and
\begin{align}
\rho_{ab}= \rho _{0}\left( x^{b}-x^{a}\right) \,,
D_{i}^{ab} =x^{b}J_{i}^{a}-x^{a}J_{i}^{b}\,.
\end{align}

To derive the effective theory for quantum fluctuations or the gapless
Goldstone modes, we expand the fields around a selected classical configuration
$\phi _{a}\left( \mathbf{x},t\right) =\sqrt{\rho _{0}+\rho _{a}\left(
\mathbf{x},t\right) }e^{i\theta _{a}\left( \mathbf{x},t\right) }$ where $%
\rho _{a}$ and $\theta _{a}$ denote density and phase fluctuations
respectively. Up to the second order, we have%
\begin{align}
\mathcal{L}& =\sum_{a=1}^{d}-\rho _{a}\partial _{t}\theta _{a}-\rho
_{0}\left( \partial _{a}\theta _{a}\right) ^{2}-\frac{1}{4\rho _{0}}\left(
\partial _{a}\rho _{a}\right) ^{2}-\frac{g}{2}\rho _{a}^{2}  \notag \\
& -\sum_{a,b}\frac{1}{2}K _{ab}\rho _{0}^{2}\left( \partial _{a}\theta
_{b}+\partial _{b}\theta _{a}\right) ^{2}\nonumber\\
&-\frac{1}{8}\sum_{a,b}K
_{ab}\left( \partial _{a}\rho _{b}+\partial _{b}\rho _{a}\right) ^{2}\,.
\label{Leff1}
\end{align}%
The density fluctuation fields $\rho _{a}$ $(a=1,\cdots,d )$ work as auxiliary fields that
are subject to the constraint equations
\begin{align}
\partial _{t}\theta _{a}=&-g\rho _{a}-\left( \frac{1}{2\rho _{0}}\partial
_{a}^{2}+\frac{1}{2}\sum_{b}K _{ab}\partial _{b}^{2}\right) \rho _{a}\nonumber\\
&-%
\frac{1}{2}\sum_{b}K _{ab}\partial _{a}\partial _{b}\rho _{b}~.
\label{Eqrhotheta}
\end{align}%
Since we are only interested in the low-energy physics, the momentum $\mathbf{k}$
has an upper bound $\left\vert \mathbf{k}\right\vert \leq 2\pi \xi _{c}^{-1}$
where the coherent length $\xi _{c}$ can be estimated as
\begin{equation}
g=4\pi ^{2}\xi _{c}^{-2}\left( \frac{1}{2\rho _{0}}+\sum_{b}K
_{ab}\right)
\label{cohcons}
\end{equation}%
as such the last two terms in Eq.~(\ref{Eqrhotheta}) can be neglected. In
the low-energy limit, we obtain the solutions $\rho _{a}=-\frac{1}{g}%
\partial _{t}\theta _{a}$. Cast it back to Eq.~(\ref{Leff1}) and we arrive
at an effective theory of quantum fluctuations by excluding the higher
derivative terms of $\theta _{a}$,
\begin{align}
\mathcal{L}=&\sum_{a}\frac{1}{2g}\left( \partial _{t}\theta _{a}\right)
^{2}-\rho _{0}\left( \partial _{a}\theta _{a}\right) ^{2}\nonumber\\
&-\sum_{a,b}\frac{1}{%
2}K _{ab}\rho _{0}^{2}\left( \partial _{a}\theta _{b}+\partial
_{b}\theta _{a}\right) ^{2} ~.  \label{efftheory}
\end{align}%
The effective theory in Eq.~(\ref{efftheory}) stays invariant under the
transformation $\theta _{a}\rightarrow \theta _{a}+\lambda _{a}+\sum_{b}\lambda
_{ab}x^{b} $ with $\lambda_{ab}=-\lambda_{ba}$. It describes $d$ gapless Goldstone modes $\theta _{a}\left( a=1,\cdots
,d\right) $ \cite{2020GoldstoneWatanabe} that have entangled motions arising from $K _{ab}$ term. 

To get a deeper insight, we concentrate ourselves on the $2$D case. Introduce the canonical
modes $\Theta _{+}\left( \mathbf{k}\right) $ and $\Theta _{-}\left( \mathbf{k%
}\right) $
\begin{align}
\Theta _{+}\left( \mathbf{k}\right) &=\cos \frac{\varphi _{\mathbf{k}}}{2}%
\theta _{1}\left( \mathbf{k}\right) +\sin \frac{\varphi _{\mathbf{k}}}{2}%
\theta _{2}\left( \mathbf{k}\right) \,,\\
\Theta _{-}\left( \mathbf{k}\right) &=-\sin \frac{\varphi _{\mathbf{k}}}{2}%
\theta _{1}\left( \mathbf{k}\right) +\cos \frac{\varphi _{\mathbf{k}}}{2}%
\theta _{2}\left( \mathbf{k}\right)\,,
\end{align}%
where $\tan \frac{\varphi _{\mathbf{k}}}{2}=\frac{2K \rho _{0}k_{1}k_{2}%
}{\Delta \left( \mathbf{k}\right)^2 }$ with denoting $K _{12}=K $ and
their dispersion relations take the form as %
$
\epsilon _{\pm }\left( \mathbf{k}\right) =\sqrt{g\rho _{0}\left[ \left( 1+K
\rho _{0}\right) \mathbf k^2\pm \Delta \left( \mathbf{k%
}\right) \right]}$, %
which is identical to Eq.~\eqref{BdGdispersion1}.
%Figure.~\ref{Fig:spectrum} depicts dispersions of the two branches of gapless modes.

Stability of a superfluid phase is determined by the long-distance behavor of the correlator of the order parameter 
under the influence of by quantum fluctuations,
\begin{align}
& \langle \text{GS}|\Phi _{a}^{\dag }\left( \mathbf{x}\right) \Phi _{b}\left(
\mathbf{0}\right) |\text{GS}\rangle _{\theta _{a}}^{\beta _{ab}}  \notag \\
=&\rho
_{0}\exp [i\sum_{c}(\beta _{bc}-\beta_{ac})x^{c}]\left\langle e^{-i\Theta
_{a }( \mathbf{x}) }e^{i\Theta _{b }( \mathbf{0})
}\right\rangle~. 
\end{align}%
We need to  calculate equal-time correlators of the canonical modes $\left\langle e^{-i\Theta
_{\pm }\left( \mathbf{x}\right) }e^{i\Theta _{\pm }\left( \mathbf{0}\right)
}\right\rangle= e^{-\frac{1}{2}\langle [\Theta_\pm(\mathbf x) - \Theta_\pm(\mathbf 0)]^2\rangle}$. Explicitly, in two spatial dimensions, we have 
%\begin{align}
%&\left\ rr1}
%\end{align}%
\begin{align}
&\left\langle \Theta _{\pm }\left( \mathbf{x}\right) \Theta _{\pm }( \mathbf 
0) \right\rangle  \notag \\
=&\int \frac{ \mathrm d\omega \mathrm d^{2}k}{\left( 2\pi \right) ^{3}}\frac{e^{i\mathbf{%
k\cdot x}}}{\omega ^{2}-\omega _{\pm }\left( \mathbf{k}\right) }  \notag \\
=& \int \frac{\mathrm d k\mathrm d\theta}{(2\pi)^2}%
\frac{e^{ik|\mathbf x|\cos \theta }}{\sqrt{g\rho _{0}\left[ \left( 1+K \rho _{0}\right) \pm \bar{%
\Delta}\left( \theta \right) \right] }} \notag \\
< & \int \frac{\mathrm d k\mathrm d\theta}{(2\pi)^2}\frac{e^{i k|\mathbf x|\cos\theta}}{\sqrt{g\rho _{0}c_{\pm}}}
=\frac{1}{2\pi |\mathbf x|}\frac{1}{\sqrt{g\rho _{0}c_{\pm}}}\,,
 \label{thetacorr}
\end{align}%
%for $\vert ^{2}\theta -\sin ^{2}\theta ) ^{2}+4K
%^{2}\rho _{0}^{2}\cos ^{2}\theta \sin ^{2}\theta }$ 
where $\bar{\Delta}( \theta ) =\sqrt{(1-2K \rho_0)\cos^2 2\theta+K^2\rho_0^2 }$ and 
$c_\pm$ denotes the minimum value of $(1+K\rho_0)\pm\bar{\Delta}( \theta )$. 
 %checked by numerics.
At the long distance $|\mathbf x|\rightarrow\infty$, the correlator $\left\langle \Theta _{\pm }\left( \mathbf{x}\right) \Theta _{\pm }(\mathbf 
0) \right\rangle$ vanishes. Thus, $\left\langle \Phi
_{a}^{\dag }\left( \mathbf{x}\right) \Phi _{b}\left( \mathbf{0}\right)
\right\rangle = \rho
_{0}\exp [i\sum_{c}(\beta _{bc}-\beta _{ac})x^{c}]$ has a finite value modulated by a plane wave. 
  It confirms  a true long-range order  that survives against quantum fluctuations when both the two lineons condensate simultaneously
  in zero temperature. Since quantum fluctuations are weaker in higher dimensions, a fractonic superfluid phase $d\mathsf{SF}^1$ stays stable in two spatial dimensions $d=2$ and higher $d>2$.  
 In Appendix~\ref{App:FlavorDep}, the component-dependent Mexican-hat potential in $2$D is considered, which leads to condensation
 of  only one component of lineons.   The uncondensed component gets released from mobility restriction
 and the Goldstone mode behaves still as a subdimensional particle. Thus, the superfluid there obeys an algebraic order.   

\section{Superfluid vortices of $2\mathsf{SF}^1$ }\label{Sec:Topovortex}

Besides the gapless Goldstone modes and gapped roton modes, thermal vortices
 are fundamental to a superfluid phase as an effect of compactness of phase
fields $\theta _{a}$. The existence of symmetry in Eq.~(\ref{momentsymm})
admits a complicated structure in $2\mathsf{SF}^1$. We present two guiding statements on
construction of point thermal vortices in $2$D and then give the two types of vortices in $2\mathsf{SF}^1$.

\subsection{Two statements on construction}
A superfluid vortex is an excitation as a consequence of compactness of a phase field
 and mathematically one can represent compactness by a multi-valued
function. Given a phase field $\theta_a$ with component $a$, we can always decompose it as
 $\theta _{a}( \mathbf{x}) =\theta _{a}^{v}( \mathbf{x})
+\theta _{a}^{s}\left( \mathbf{x}\right) $ where $\theta _{a}^{s}(
\mathbf{x}) $ denotes the smooth component.
 In general, the multi-valued component $\theta_{a}^{v}\left( \mathbf{x}\right) $ can be formulated as
\begin{equation}
\theta _{a}^{v}( \mathbf{x}) =f_{a}( \mathbf{x})
\varphi ( \mathbf{x})\,,  \label{thetavortex}
\end{equation}%
where  $\varphi ( \mathbf{x})$ defined $\mathrm{mod}$ $2\pi $
is the angle of site $\mathbf x$ relative to vortex core and $f_a(\mathbf x) $ is a single-valued function. Eq.~(\ref{thetavortex}) sets an equivalent
relation $\theta _{a}^{v}\left( \mathbf{x}\right) \sim \theta _{a}^{v}\left(\mathbf{x}\right) +2\pi f_{a}\left( \mathbf{x}\right) $. Subtly, $%
f_{a}( \mathbf{x}) $ should be understood under a lattice
regularization to protect single-valuedness of field $\phi_a$, where spatial coordinates $\mathbf{x}$ are regarded as $%
\mathbf{x}=(x^1,x^2)=\mathbf{n}a$ with $\mathbf{n}=\left( n_{1},n_{2}\right) $ being a pair of
integers and $a$ being the lattice constant. 
The equivalence relation in Eq.~(\ref%
{thetavortex}) resembles a gauge freedom. Whether we start with $\theta
_{a}^{v}\left( \mathbf{x}\right) $ or $\theta _{a}^{v}\left( \mathbf{x}%
\right) +2\pi f_{a}\left( \mathbf{x}\right) $ should cause no physical effects. 
Therefore, we arrive at  Statement~\ref{statement-I} below:
\begin{statement}
 The physical Hamiltonian density  should be single-valued even in the presence of multi-valued vortex configurations. 
\label{statement-I}
\end{statement}
  Statement~\ref{statement-I}  clarifies that the Hamiltonian density $\mathcal H[\theta_a(\mathbf x)]$ is invariant when $\theta_a(\mathbf x)$ is shifted by $2\pi f_a(\mathbf x)$, $\mathcal{H}[\theta_a(\mathbf x)+2\pi f_a(\mathbf x)]= \mathcal H[\theta_a(\mathbf x)]$, which determines the most singular part of a vortex. 
 Take a conventional superfluid $2\mathsf{SF}^2$ as an example with Hamiltonian density $\mathcal{H}=\frac{1}{2}[(\partial_1\theta(\mathbf x))^2+ (\partial_2\theta(\mathbf x))^2]$. With the assumption $\theta^v(\mathbf x)=f(\mathbf x)\varphi(\mathbf x)$, the constraint imposed by Statement~\ref{statement-I} on shifting $\theta(\mathbf x)$ 
by $2\pi f(\mathbf x)$ gives the equations $\partial_1 f(\mathbf x)=0, \partial_2 f(\mathbf x)=0$ towards which we have the solution $f(\mathbf{x})=\ell$ with $\ell\in \mathbb Z$. Thus we recover vortex configurations in a conventional superfluid. 

The second statement to be introduced below controls the smooth component after we obtain the multi-valued component from  Statement~\ref{statement-I}. 
A higher-rank symmetry group contains not only conventional $U(1)$ charges that induce a global $U(1)$ phase shift, but also charges that generate 
a  phase shift depending on local coordinates. For convenience, we call  conventional $U(1)$ charges as rank-$0$ while the others are higher-rank charges. Statement~\ref{statement-2} below establishes the relations between higher-rank and rank-$0$ charges: 
\begin{statement}
The action of a higher-rank symmetry group on some bound states of operators charged in the higher-rank symmetry group is equivalent to an action of a global $U(1)$ symmetry with appropriate rank-$0$ charges. 
\label{statement-2}
\end{statement}
Statement~\ref{statement-2} allows us to construct a set of bound states 
such that  the higher-rank group only induces a global  phase shift. Explicitly, given vortices carrying higher-rank charges, Statement~\ref{statement-2} claims that some bound state of these vortices is proportional to $\varphi(\mathbf x)$ as a conventional vortex, that is, the smooth component vanishes.
Thus, the essence is to find the structures of bound states which  are significantly determined by relations between higher-rank charges and rank-$0$ charges. 
For example, we consider a higher-rank symmetry \cite{2020PhRvR2b3267Y} which shifts $\theta(\mathbf x)$ by $\theta(\mathbf x)\rightarrow \theta(\mathbf x)+ \lambda+\sum_a\lambda_a x^a$. Then the group action on bound states like $\hat{\mathcal{O}}_{\mathbf d}=e^{-i\theta(\mathbf x)}e^{i\theta(\mathbf x-\mathbf d)}$ with a constant vector $\mathbf d$ generates a pure global phase, $\hat{\mathcal{O}}_{\mathbf d}\rightarrow \hat{\mathcal{O}}_{\mathbf d}e^{-i\sum_a \lambda_a d^a}$. Thus, on these bound states, the higher-rank group is equivalent to group $U(1)$  and we shall expect that $\hat{\mathcal{O}}_{\mathbf d}$ takes the form of a conventional vortex whose smooth part can be set to vanish. In the appendix~\ref{App:app}, we present a detailed derivation on vortices in a higher-rank superfluid phase in Ref.~\onlinecite{2020PhRvR2b3267Y}.

\subsection{Vortex structure}
The two statements are more generally applicable for a system of a higher-rank symmetry.  
At present, we specialize our attention to the case of $2\mathsf{SF}^1$. 
Statement~\ref{statement-I} leads to an assumption for the multi-valued component
\begin{equation}
\theta_1^v(\mathbf x)=f_1(\mathbf x)\varphi(\mathbf x)\,,  \,\theta_2^v(\mathbf x)=f_2(\mathbf x)\varphi(\mathbf x)\,,
\label{assumf12}
\end{equation}
and $f_{1,2}(\mathbf x)$  should satisfy the following equations:
\begin{equation}
\partial_1 f_1(\mathbf x)=0\,,\, \partial_2 f_2(\mathbf x)=0\,,\, \partial_2 f_1(\mathbf x)+\partial_1 f_2(\mathbf x)=0.
\end{equation}
 The solutions generally can be parametrized by three parameters,
 \begin{equation}
 f_1(\mathbf x)= p x^2+ \ell_1, f_2(\mathbf x)=-p x^1+\ell_2\,.
 \end{equation}
 Here, $p$ has the dimension $[x]^{-1}$, which we dub  a dipole charge, while $\ell_1$ and $\ell_2$ are dimensionless. Under the lattice regularization, $pa$ and $\ell_1,\ell_2$ are all integers. 
 
The parameters $\ell_1$ and $\ell_2$ describe conventional vortices and  they are interpreted as the winding numbers. To obtain vortices carrying a dipole charge $p$, we apply Statement~\ref{statement-2}. 
The essence of Statement~\ref{statement-2} is to recognize bound states. 
In $2$D, the higher-rank group $\mathcal G$ is parametrized by $\lambda_1,\lambda_2$ and $\lambda_{12}$. We denote the group element with $\lambda_1=\lambda_2=0,\lambda_{12}=1$ as $G_e$. 
Given  a vortex  operator $\hat{\mathcal{O}}_1(\mathbf x)=e^{i\theta_1(\mathbf x)}$ with the charge $-q_{12}^{}$ , which is transformed by $G_e$ in $\mathcal G$ as  $\hat{\mathcal{O}}_{1}\left( \mathbf{x}\right) \rightarrow \hat{\mathcal{O}}_{1}\left( \mathbf{x}\right) e^{i q_{12}^{} x^{2}}$, then a `particle-hole' bound state
$\hat{\mathcal O}_1^\dagger(\mathbf x)\hat{\mathcal O}_1(\mathbf{x}-\mathbf d)$ with a constant vector $\mathbf d=(0,d)$ is transformed by $G_e$ as 
\begin{equation}
\hat{\mathcal O}_1^\dagger(\mathbf x)\hat{\mathcal O}_1(\mathbf{x}-\mathbf d)\rightarrow \hat{\mathcal O}_1^\dagger(\mathbf x)\hat{\mathcal O}_1(\mathbf{x}-\mathbf d)e^{-iq_{12}^{} d} 
\label{compositedipole}
\end{equation}
If the particle-hole bound state is attributed with a $U(1)$ charge $q_{12}^{}d$, we can find the action in Eq.~\eqref{compositedipole}
can be re-explained as action of $U(1)$ symmetry. Statement~\ref{statement-2} asserts that the bound state reduces to a conventional vortex, which imposes a  constraint $\theta _{1}\left( \mathbf{x%
}\right) -\theta _{1}\left( \mathbf{x}-\mathbf{d}\right) =d\partial_2\theta _{1}\left( \mathbf{x}\right)= q_{12}^{} d \varphi(\mathbf x) $ for small $\mathbf{d}$. We have
\begin{align}
pd\varphi(\mathbf x)-q_{12}^{}d\varphi(\mathbf x) &=0~,  \label{vor1p}\\
\partial _{2}\theta _{1}^{s}\left( \mathbf{x}\right) +f_{1}\left( \mathbf{x}%
\right) \partial _{2}\varphi \left( \mathbf{x}\right) &=0~.
\end{align}%
Eq.~\eqref{vor1p} shows $p=q_{12}^{}$ which indicates the dipole charge $p$ in $f_1(\mathbf x)$ represents a higher-rank charge of group $\mathcal G$.
 In fact, the bound state with $\mathbf d=(d,0)$ is invariant under $G_e$ and thus it requires  $\hat{\mathcal O}_1^\dagger(\mathbf x)\hat{\mathcal O}_1(\mathbf{x}-\mathbf d) $ to be single-valued, which is satisfied since $f_1(\mathbf x)$ is independent of $x^1$. 
We consider vortex operator object $\hat{\mathcal{O}}_2(\mathbf x)=e^{i\theta_2}$ carrying a charge $q_{12}^{}$ with a transformation by $G_e$ as $\hat{\mathcal{O}}_{2}\left( \mathbf{x}\right) \rightarrow \hat{\mathcal{O}}_{2}\left( \mathbf{x}\right) e^{-i q_{12}^{}x^{1}}$. Then $G_e$ induces a global  phase shift on the bound state $\hat{\mathcal{O}}_2(\mathbf x)^\dagger \hat{\mathcal{O}}_2(\mathbf x-\mathbf d)\rightarrow \hat{\mathcal{O}}_2(\mathbf x)^\dagger \hat{\mathcal{O}}_2(\mathbf x-\mathbf d)e^{iq_{12}^{}d}$ with $\mathbf d=(d,0)$. Thus, we are allowed to re-interpret action of $\mathcal G$ on a bound state $\hat{\mathcal{O}}_2(\mathbf x)^\dagger \hat{\mathcal{O}}_2(\mathbf x-\mathbf d)$ as an action of  $U(1)$ group on a charged $-q_{12}^{}d$ operator. Therefore, according to Statement~\ref{statement-2}, we have $\theta_{2}( \mathbf{x})-\theta_{2}( \mathbf{x}-\mathbf d)=-q_{12}d\varphi(\mathbf x)$ for small $d$. Equivalently, we have
\begin{align}
-pd\varphi(\mathbf x)+q_{12}d\varphi(\mathbf x)&=0~,\\
\partial _{1}\theta _{2}^{s}\left( \mathbf{x}\right) +f_{2}\left( \mathbf{x}%
\right) \partial _{1}\varphi \left( \mathbf{x}\right) &=0~.
\end{align}%
Here the dipole charge denotes a higher-rank charge of group $\mathcal G$.
%
 %\right) \partial _{1}\varphi \left( \mathbf{x}\right) =0~.
%\end{equation}%
Thus, we can obtain two types of vortices. The first one is the  conventional
vortex characterized by winding numbers $\ell_1,\ell_2$
\begin{equation}
\theta_1(\mathbf x)=\ell_1 \varphi(\mathbf x), \theta_2(\mathbf x)=\ell_2 \varphi(\mathbf x)~.
\end{equation}
And the second one with a configuration
%\begin{align}
  %\end{align}%
\begin{align}
\theta _{1}\left( \mathbf{x}\right) & =-px^{1}\log |\mathbf x|+px^{2}\varphi \left( \mathbf{x}\right) \\
\theta _{2}\left( \mathbf{x}\right) & =-px^{2}\log |\mathbf x|-px^{1}\varphi \left( \mathbf{x}\right)
\end{align}%
carries a higher-rank charge $p$ of symmetry $ \mathcal{G}$.
We emphasize again that the charge $p$ should be regularized as $p=\ell a^{-1}$ $\left( \ell\in \mathbb{Z}%
\right) $ to ensure $f_{1,2}\left( \mathbf{x}\right) $ in Eq.~(\ref{assumf12}) to be integer-valued.
When we circle around the vortex core, the vortex configuration get an extra
phase $\delta \theta _{1}=2\pi px^{2}=2\pi \ell n_{2}$ and $\delta \theta
_{2}=-2\pi px=-2\pi \ell n_{1}$ with $\mathbf x=(n_1,n_2)a$, which keeps in consistence with compactness of $%
\theta _{a}$. Different from a conventional vortex, here $\partial
_{1}\theta _{2}$ and $\partial _{2}\theta _{1}$ are still multi-valued while $\partial_1\theta_1$ and
$\partial_2\theta_2$ are single-valued.

 We can define the vorticity for the dipole charge by recombination of Noether currents.
After condensation, in two spatial dimensions the Noether currents can be formulated as
\begin{equation}
\!\!\!\!\!J_{1}^{1}\!\! =2\rho _{0}\partial _{1}\theta _{1} , J_{2}^{2} \!\!=2\rho _{0}\partial _{2}\theta _{2} ,
J_{1}^{2} \!\!=2K \rho _{0}^{2}\left( \partial _{1}\theta _{2}+\partial
_{2}\theta _{1}\right).
\label{currecond}
\end{equation}
As indicated by Statement~\ref{statement-2}, a `particle-hole' bound state of vortices behaves as a vortex in $2\mathsf{SF}^2$
 and it only encodes the dipole charge. Above all, the density $\rho_\mathrm{dipole}$ of such a bound state can be written as
\begin{equation}
\rho _{\mathrm{dipole}}=\sum_{a,b=1}^{2}\frac{1}{2}\epsilon _{ab}\partial
_{a}\partial _{b}\left( \partial _{2}\theta _{1}-\partial _{1}\theta
_{2}\right)~,
\end{equation}%
where $\epsilon_{ab}$ is an  antisymmetric tensor $\epsilon_{12}=-\epsilon_{21}=1$.
Actually we have a relation
\begin{align}
\rho _{\mathrm{dipole}}&=\sum_{a,b=1}^{2}\epsilon _{ab}\partial
_{a}\partial _{b} \partial _{2}\theta _{1} = \sum_{a,b=1}^{2}-\epsilon _{ab}\partial
_{a}\partial _{b}\partial _{1}\theta
_{2}~,
\end{align}
 since $\theta_1$ and $\theta_2$ take the same dipole charge.
 Following the lesson we learnt for vortices in superfluid phase $2\mathsf{SF}^2$,
 we can construct the currents $\mathbf{J}_{\text{dipole}}$ based on
 condensed currents in Eq.~\eqref{currecond} with components
\begin{align}
\mathbf J_{\mathrm{dipole}}^{1} &=-\frac{1}{2}aK ^{-1}\rho _{0}^{-2}\partial
_{1}J_{1}^{2}+a\rho _{0}^{-1}\partial _{2}J_{1}^{1}\,, \\
\mathbf J_{\mathrm{dipole}}^{2} &=\frac{1}{2}aK ^{-1}\rho _{0}^{-2}\partial
_{2}J_{1}^{2}-a\rho _{0}^{-1}\partial _{1}J_{2}^{2} ~,
\end{align}%
where the cutoff $a$ is introduced to make up the dimension of $\mathbf{J}_\mathrm{%
dipole}$. Then the vorticity will give the dipole charge,
\begin{equation}
\ell _{\mathrm{dipole}}=\frac{1}{2\pi }\oint\limits_{C} \mathrm d\mathbf{x}\cdot
\mathbf{J}_{\mathrm{dipole}}=\ell~,
\end{equation}
where $C$ is a closed path encircling the vortex core and $p=\ell a^{-1}$.

%  number of vortex of field $\phi _{a}$.

\section{Concluding Remark}
\label{Sec::concl}
As a series of the work \cite{2020PhRvR2b3267Y}, we have further explored more possibilities of exotic states of matter formed by particles with restricted mobility. We have discussed a fractonic superfluid phase $d\mathsf{SF}^1 $ in a microscopic model 
by condensing subdimensional particles. This model is invariant under a higher-rank symmetry such that 
its fundamental particles are lineons.
We use the HFB mean-field theory to derive  a set of highly non-linear GP equation and a BdG Hamiltonian which characterize the condensed and the norm components respectively.  In the framework of a continuous field theory, we construct macroscopic degeneracies of ground states with finite momentum and
derive an effective theory for gapless Goldstone modes. At zero temperature, a phase $d\mathsf{SF}^1 $ stays stable in two spatial dimensions and higher.
We emphasize two guiding statements to construct vortex excitations in two spatial dimensions.
Explicitly, there are two types of vortices in $2\mathsf{SF}^1$. Besides conventional vortices, the other type carries a dipole charge.  The two guiding statements are more generally applicable [See Appendix~\ref{App:app}].

Towards a complete understanding on a fractonic superfluid phase, we have to deal with more questions. Tightly related to the present paper, vortex excitations form a hierarchy which is dominated by the two statements, and then interactions between vortices and BKT transitions should also inherit such a hierarchy. A natural question is to investigate a superfluid phase by condensing other spatially extended excitations \cite{ye19a}.  
On the other hand, by viewing the model in Eq.~(\ref{Ham}) as coupled  Luttinger liquids, we can derive 
an effective theory by  bosonization. Then, it is straightforward to obtain universal properties of fractonic superfluids such as   conductivity and general thermodynamic properties \cite{2002PhRvB66e4526P, 2020arXiv201015148S}.
In three spatial dimensions, more exotic vortex line excitations can be excited and their construction needs further investigation.   If we condense   these defects to recover the symmetry as the scheme to construct a symmetry protect topological phase \cite{Chen:2014aa,PhysRevB.93.115136,YeGu2015,bti2,yp18prl}, what phase can be obtained? 
Besides, what is the universal class of the phase transition between a high-rank superfluid phase and a normal state?
Experimentally, we expect that 
the Hamiltonian in Eq.~\eqref{Ham}  can be realized in the cold atomic gas subjected to an optical lattice 
by tuning a two-particle states \cite{ExperimentFisher2005}, 
which opens a new horizon to search exotic phases of matter.

\acknowledgements 
We thank Yuxuan Wang, Zhi Wang, Wen Huang and Jian- Hua Jiang for their useful discussions. This work was supported in part by the Sun Yat-sen University startup grant, Guangdong-Shenzhen Regional Joint Fund (Key Program) of Guangdong Natural Science Foundation (Grant No. 2020B1515120100), and National Natural Science Foundation of China (NSFC) (Grants No. 11847608 and No. 12074438).

%\newpage
\appendix
\section{Component-dependent potential}\label{App:FlavorDep}
In the main text, a Mexican-hat potential is chosen to be component-indepdent and the two components of lineons are condensed simultaneously.  
Here, we briefly discuss a component-dependent Mexican-hat potential in two spatial dimensions with only one of the two components is condensated.
In the framework of coherent-path integral representation, we introduce a Mexican-hat potential only for $\phi_2$, 
%\begin{equation}
%V=-\mu_1 |\phi_1|^2+ \frac{g_1}{2}|\phi_1|^4-\mu_2  |\phi_2|^2+ \frac{g_2}{2}|\phi_2|^4
%\end{equation}
\begin{equation}
V=-\mu_2  |\phi_2|^2+ \frac{g_2}{2}|\phi_2|^4~,
\label{App:Vphi2}
\end{equation}
with $g_2>0$.
When $\mu_2<0$, $\phi_2$ is in an insulating state. When $\mu_2>0$,   $\phi_2$ picks up a finite particle density,
$\rho_{20}\equiv \frac{\mu_2}{g_2}$ to minimize the potential in Eq.~\ref{App:Vphi2}. In the classical level, the configurations of $\phi_2$ take the 
form as
\begin{equation}
\phi_2^\mathrm{cl}=\sqrt{\rho_{20}}e^{i\theta_2 +i\beta_{12}x^2}~.
\label{App:phi2}
\end{equation}
We can expand the field $\phi_2$ around $\phi_2^\mathrm{cl}$ in Eq.~\eqref{App:phi2}. The obtained Hamiltonian 
explicitly depends on the parameter $\beta_{12}$. This dependence can be removed by a
transformation
\begin{equation}
\phi_1(\mathbf x) \rightarrow \phi_1(\mathbf x) e^{-i\beta_{12}x^1}~.
\end{equation}
Follow the procedure in Sec.~\ref{section_eft111} and up to second order, we obtain an 
effective theory,
\begin{align}
\mathcal{L}[\phi_1]&= i\phi_1^*\partial_t\phi_1^{}-|\partial_1\phi_1|^2-K\rho_{20}|\partial_2\phi_1|^2  \label{App:L_phi1}\\
\mathcal{L}[\theta_2] & =\frac{1}{2g_2}(\partial_t \theta_2)^2-\rho_{20}(\partial_2\theta_2)^2 
\end{align}
where $K=K_{12}$.
The  symmetry of $\phi_1$ in Eq.~(\ref{App:L_phi1}) reduces to a conventional particle conservation $U(1)$ symmetry after $\phi_2$ condensation,
and thus the field $\phi_1$ gets  liberated from the mobility restriction due to the last term in Eq.~(\ref{App:L_phi1}).
However, the Goldstone mode $\theta_2$ still lacks full mobility and can only propagate in one direction.
The stability can be inferred from the long-range behaviours of order parameter correlator $\langle e^{i\theta_2(\mathbf x)}e^{-i\theta_2(\mathbf 0)}\rangle =e^{-\frac{1}{2}\langle (\theta_2(\mathbf x)-\theta_2(\mathbf 0))^2\rangle}$. 
which is determined by 
\begin{equation}
\langle \theta_2(\mathbf x) \theta_2(\mathbf 0)\rangle =\delta( x^1) \int \frac{\mathrm d\omega \mathrm  dk_2}{(2\pi)^3} e^{-ik_2 x^2}\frac{g_2}{\omega^2- \mu_2 k_2^2}
\label{App:Cor}
\end{equation}
If $x^1\not = 0$, the correlator in Eq.~\eqref{App:Cor} vanishes as a natural consequence of mobility constraint. When $x^1=0$, 
we have
\begin{equation}
\langle \theta_2(\mathbf x) \theta_2(\mathbf 0)\rangle = -\frac{2\pi}{\xi}\frac{\pi g_2}{\sqrt{ \mu_2}} \log\frac{4 e^{-\gamma}(x^2)^2}{\mu_2}
\end{equation}
such that $\langle e^{i\theta_2(\mathbf x)}e^{-i\theta_2(\mathbf 0)}\rangle$ decays in a power-law pattern where $\xi$ is the coherent length 
and $\gamma$ is  the Euler constant. Therefore, the superfluid phase here is algebraically ordered.

\section{Application of the two statements}
\label{App:app}
In the mainbody, we put forward  two statements to construct vortex excitations and 
indicate they are generally applicable. Here, we apply them to a fracton model that is considered in Ref.~\cite{2020PhRvR2b3267Y}.
The Hamiltonian density for the Goldstone modes reads
\begin{equation}
\mathcal H=\frac{1}{2}[ (\partial_1^2\theta)^2+ 2(\partial_1\partial_2\theta)^2+(\partial_2^2\theta)^2]~.
\label{app:mdl}
\end{equation}
The vortex excitations arise when $\theta$ is multi-valued. 
We decompose vortex field as $\theta=\theta^v+\theta^s$, where the multi-valued component can be formulated as 
\begin{equation}
\theta^v(\mathbf x)=f(\mathbf x)\varphi(\mathbf x)~,
\label{app:thetav}
\end{equation}
with $\varphi(\mathbf x)$ is the angle of site $\mathbf x=(x^1,x^2)$ relative to vortex core. 
Statement~\ref{statement-I} requires Hamiltonian density $\mathcal H$ is single-valued for the field configuration in Eq.~\eqref{app:thetav}, which induces the restrictions on $f(\mathbf x)$,
\begin{equation}
\partial_1 f(\mathbf x)=0,\quad \partial_2 f(\mathbf{x})=0~.
\end{equation}
It is easy to find the solutions, 
\begin{equation}
f(\mathbf x)=\ell~,
\label{app:fell}
\end{equation}
or
\begin{equation}
f(\mathbf x)=p_1 x^1, \quad f(\mathbf x)=p_2 x^2~.
\label{app:fp}
\end{equation}
The first solution in Eq.~\eqref{app:fell} marks a conventional vortex solution with 
$ \theta=\ell \varphi(\mathbf x)$ where $\ell\in \mathbb Z$ represents the winding number.
The second solution in Eq.~\eqref{app:fp} takes charges $p_1,p_2$ with dimension $[x^{-1}] $. The charges are expect to be 
quantized as a momentum. The full solution should take into consideration the smooth component $\theta^s(\mathbf x)$,
which is instructed by Statement~\ref{statement-2}. For $f(\mathbf x)=p_1x^1$, we  consider a bound state of a vortex-anti-vortex pair with distance $\mathbf d=(d,0)$, on which the vortex bound state reduces to a conventional vortex with winding number $dq_1$,  i.e.
\begin{equation}
d\partial_1[p_1 x^1\varphi(\mathbf x)+\theta^s(\mathbf x)]=dq_1\varphi(\mathbf x)
\end{equation}
So $\theta^s(\mathbf x)=q_1x^2\log|\mathbf x|$. Similarly, consider a bound state of a  vortex-anti-vortex pair with  distance  $\mathbf d=(0,d)$ and we can obtain the smooth component for $f(\mathbf x)=-q_2x\log|\mathbf x|$.
In summary, vortices for the model in Eq.~(\ref{app:mdl}) have the configurations as 
\begin{align}
\theta &=\ell \varphi(\mathbf x)~, \\
\theta & = p_1 x^1\varphi(\mathbf x)+p_1 y\log|\mathbf x|~,\\
\theta & = p_2 x^2\varphi(\mathbf x)-p_2  x\log|\mathbf x|~.
\end{align}

%\bibliography{top}
%merlin.mbs apsrev4-1.bst 2010-07-25 4.21a (PWD, AO, DPC) hacked
%Control: key (0)
%Control: author (0) dotless jnrlst
%Control: editor formatted (1) identically to author
%Control: production of article title (0) allowed
%Control: page (1) range
%Control: year (0) verbatim
%Control: production of eprint (0) enabled
%

\end{document}